\documentstyle[prl,aps]{revtex}
\begin{document}
\draft


 \title{Quantum Phase Transition in Skyrmion Lattice}

 \author{Yu.V. Nazarov and A.V. Khaetskii \cite{address1}}

 \address{ Faculty of Applied Physics and DIMES, Delft University of Technology,\\
Lorentzweg 1, 2628 CJ Delft, The Netherlands }

\maketitle

\begin{abstract}
We investigate the ground state of  2D electron gas in Quantum Hall regime
at the filling factor slightly deviating from unity, that can be viewed
as a sparse lattice of skyrmions. We have found that in the low density limit
skyrmions are bound in pairs, those  forming the actual lattice.
We have shown  that at further density increase the lattice undergoes 
a quantum phase transition, an analogue of superconducting phase transition
in Josephson junction arrays.    
\end{abstract}

Skyrmion \cite{Skyrme} has been a favorite model particle of field 
theorists for decades. Only recently it has been comprehended 
that the low-lying charged excitations of the Integer
Quantum Hall ground state may be skyrmions(\cite{Sondhi}, see 
\cite{MacReview} for review).
This interesting development has prompted intensive experimental
studies \cite{Barrett,English} that  have proved 
these exotic particles to be real indeed.
A skyrmion can be viewed as a topologically non-trivial
distortion of the spins of {\it many} ground state electrons.
That makes its spin much bigger than unity, 
although the skyrmion bears  a unitary charge.
The size of the skyrmions is controlled by a parameter $\tilde g$, the
ratio of Zeeman energy to exchange energy per particle. 
The smaller the  $\tilde g$ is, the bigger is the skyrmion.

If the  filling factor slightly deviates from unity, the low-lying
charged excitations must appear at the background of the Quantum Hall
state to compensate for excess charge.
Therefore, the deviation of the filling factor sets the skyrmion concentration
and the many-skyrmion ground state can be easily made
by changing either magnetic field or electron density of 2D gas.
The two studies of many-skyrmion ground state at 
finite $\tilde g$ have been
recently reported. \cite{Brey,Green}
Despite the different methods applied and the different results obtained, 
it has been assumed in both studies that the skyrmions form 
a plain regular lattice. Also, in both studies  no quantum effects have been taken
into account. 

Our results demonstrate that the formation and development of the
multi-skyrmion state is more complicated and interesting than
it has been assumed previously. We have shown
that at low concentrations the skyrmions always appear in pairs.
These pairs form a triangular Wigner lattice. 
With increasing concentration the distance between pairs decreases,
and the interaction between the pairs strengthens. This
leads to a quantum transition in skyrmion lattice. 
Below the transition point, the z-component
of spin of each pair is well-defined. We call  the variable
that is canonically conjugated to z-component of spin the {\it phase}. 
Above the transition
point, it is this phase that is well-defined.

We begin with the energy functional that describes  spin textures
in the limit of small $\tilde g$ (see e.g. \cite{Sondhi,Green}). 
\begin{eqnarray}
E= E_{stiff} + E_Z + E_{Coulomb} \nonumber \\
= \int d^2 r \left [ \frac{\rho _{s}}{2} |\partial _{\mu } {\bf m}|^2  + 
\frac{|g|\mu _B B n_L}{2} (1+m_z) \right ]
+ \frac{1}{2} \int d^2 r \int d^2 r' q({\bf r}) \frac{e^2}{\epsilon|{\bf r} -
{\bf r'}|} q({\bf r'}).
\label{1}
\end{eqnarray}
It contains three terms that  describe respectively spin stiffness,
Zeeman energy, and electrostatic energy.
It depends on ${\bf m }({\bf r}) $, a unitary vector characterizing  local spin
density.
We adopt the microscopic expression \cite{BychOld}
for the spin stifness at  filling factor 1: 
$\rho _{s} = (1/16\sqrt {2\pi}) (e^2 / \epsilon \lambda )$, $\lambda$ being
the magnetic length. In the Zeeman term, $g$ stands for the electron g-factor, 
and $\mu _B$ is the Bohr magneton. 
 In the electrostatic term, 
$q({\bf r})$ presents  the 
deviation of the electron density from the uniform background 
density $n_L = 1/2\pi \lambda ^2$, $q$  being related to 
the density of topological charge \cite{Sondhi}:
\begin{equation}
q = -\frac{1}{8\pi}\epsilon _{\mu \nu} ({\bf m}[\partial _{\mu }{\bf m} \times 
\partial _{\nu }{\bf m} ]).
\label{2}
\end{equation}
The precise definition of $\tilde g$ reads $\tilde g \equiv g\mu _B B/E_{ex}, 
E_{ex} \equiv e^2/\epsilon \lambda $. 

The stiffness term dominates the energy and must be minimized first of all.
It has been shown by Belavin and Polyakov \cite{Polyakov}
that the textures minimizing the stiffness term are highly
degenerate.  
They parametrized the spin density as follows:
\begin{equation}
W= \frac{m_x +im_y }{1-m_z}, \quad m_x +im_y= \frac{2W}{1+ |W|^2}, \quad m_z = 
\frac{|W|^2 -1}{|W|^2 +1}.
\label{3}
\end{equation}
It turns out that all the textures corresponding
to any {\it analytical} function $W(z \equiv x+iy)$
with $N$ poles possess the same energy $N \sqrt{\pi /32}E_{ex}$. 
Neither pole (skyrmion
positions) nor pole residuals (skyrmion radii) are fixed. Those are
determined by the interplay of the weaker interactions:
Zeeman and electrostatic energy.

Let us start with a single skyrmion.\cite{Sondhi} We take $W(z)= a/z$, $a$
being the skyrmion radius to be evaluated. The Coulomb energy
is given by $E_Q = (3\pi ^2 /64)e^2/\epsilon a$. The integral that gives
Zeeman energy appears to diverge logarithmically at long distances
from the core, since $x,y$ spin components decrease very slowly
with increasing  distance. A slightly improved calculation     
shows that the divergency is to be cut at $r_s \simeq \lambda \tilde g^{- 1/2}$, and Zeeman energy
reads $E_Z = g\mu _B B  (a^2 /\lambda ^2) \ln (r_s/a)$ provided $a \ll r_s$. 
The minimization
of the total energy with respect to $a$ yields 
$a \simeq \lambda (ln(1/\tilde g)\tilde g)^{- 1/3}$,
so that $a \ll r_s$ indeed.

Let us  consider now {\it two} skyrmions with {\it opposite} phases,
\begin{equation}
W(z)= \frac{a}{z+ R} - \frac{a}{z-R}, 
\end{equation}
 the minus sign before the second term  accounting for that phase shift.
We see that the $x,y$ spin components are effectively quenched at
$|z|\gg R$, that lowers Zeeman energy. For well-separated skyrmions ($a \ll R$)
the logarithmical divergency is cut at distances of the order of $R$ and Zeeman energy reads
$2 g\mu _B B  (a^2 /\lambda ^2) \ln (R/a)$. This results in an attractive
force between skyrmions $\propto 1/R$. At long distances that always
prevails Coulomb repulsion, so that the skyrmions are bound to form pairs.
To find how they are exactly bound, we have studied the question numerically.
Fig. 1 shows our results for the two skyrmion energy as function of their separation
$R$, at optimal $a$. It appears that the minimum is achieved at zero
separation, corresponding to $W(z)=b^2/z^2$. Two skyrmions merge as it sketched
in Fig. 1. The optimal value of $|b|$ is
$b_0 \simeq  \lambda / \tilde g ^{1/3}$. The skyrmion pair bears big spin,
$S_0 \simeq 0.78 / \tilde g^{2/3}$. We checked numerically that skyrmion
attraction saturates, that is, there is no further bounding of pairs.

Therefore we have shown that the charged excitations with lowest energy
are eventually skyrmion pairs bearing charge $\pm 2 e$. We proceed now
with analysing many-skyrmion ground state.\cite{Super} Let us compare
now  Zeeman and  Coulomb contributions to the interaction between pairs.
The latter appears to prevail provided the separation between pairs exceeds their
equilibrium radius $b_0$. Therefore, the pairs behave very much like point-like
charges and form a triangular Wigner lattice to minimize their Coulomb energy.
Such an arrangement persist up to relatively high skyrmion densities 
$n_{sky}/n_{el} \simeq (b_0/\lambda)^{-2} \simeq \tilde g^{2/3}$ where they begin to overlap. 
Most probably, the overlap
would depair skyrmions and one of the
single-skyrmion lattices proposed would be realized at higher densities.
In the present paper, we restrict our attention to lower densities where
the pairs certainly form the triangular lattice.

To reveal the interesting physics that persists even in this limiting case,
let us note that the two-skyrmion solution found is infinitely degenerate.
All the spins can be rotated about $z$-axis by an arbitrary angle yielding
the texture of the same energy. Or, alternatively, we can multiply $W(z)$
by an arbitrary phase factor $\exp (i\phi)$. We will call  the
variable parametrizing all the degenerate textures the {\it phase}. This degeneracy is
not physical and arises from the fact that we treat the pair, which is
a quantum particle, as a classical object.
 
Let us consider quantization of an isolated pair. We note that the
shift in the phase space, $\phi \rightarrow \phi +\theta$, is just
a rotation in spin space about $z$-axis being represented by the
quantummechanical operator $\exp( -i \theta \hat S_z)$. Here $\hat S_z$ stands 
for the total spin of the skyrmion. Expanding near $\theta=0$, we find that $S_z = -i \partial/\partial \phi$.
That leads us to an important conslusion:
\begin{equation}
[ \hat \phi, \hat S_z] = i;
\end{equation}  
the phase and the total spin of the pair are conjugate variables satisfying
Heisenberg uncertainty relation. Owing to degeneracy with respect to phase, 
physical isolated skyrmion pairs do not have
a certain phase but rather possess a certain integer $z$-component of spin.
This picture has very much in common with Coulomb blockade in superconducting
island \cite{Efetov,Schon}, that does not have a certain superconducting phase
but rather a certain integer charge.  

Since the total spin of the pair is a function of its radius $b$, as well as
its energy, the spin quantization leads to the energy quantization of
pair states. The spectrum can be obtained from classical dependences $E(b)$ and
$S_z(b)$. Near the minimum energy it reads
\begin{equation} 
E= E_0 +E_c (S_z-S_0)^2, \quad E_c \equiv  
\frac{3}{4} \frac{\tilde g E_{ex}}{S_0}.
\label{8}
\end{equation}   
Here $E_0,S_0$ are optimal classical energy and spin respectively.
Let us note that $S_0$ is a continious function of $\tilde g$, to be contrasted
with integer $S_z$. If we change $\tilde g$, $S_0$ changes continuously, whereas
the optimal $S_z$ jumps between integer values. In the point of jump, energies
of two states with different $S_z$ match providing extra degeneracy.

 This is reminiscent of Coulomb blockade phenomenon in Josephson
arrays. \cite{Schon} Josephson arrays consist of superconducting
islands, the number of discrete charges in each island being
canonically conjugate to its superconducting phase. The actual state
of the array is determined by an interplay of charging energy 
and Josephson coupling.
If charging energy dominates, the array is in insulating phase.
There is a gap for charged excitations: one has to pay charging
energy in order to add or extract a discrete charge to/from the array. 
If the Josephson coupling increases, the quantum phase transition to
the superconducting phase takes place. The long-range phase ordering appears,
the gap vanishes. 

  We shall expect similar behaviour from skyrmion lattices, with discrete
charge replaced by discrete $S_z$. At very low skyrmion concentrations the pairs 
can be regarded as isolated ones
each having a well-defined spin and strongly fluctuating phase. The $x,y$ components
of the local spin density are undefined and have zero average, as it is sketched
in Fig. 2a. 
There is a "spin gap" in this "inslulating" phase: one has to pay energy of the
order of $E_c$ to increase or decrease the spin of the system by unity.
If we increase skyrmion concentration,
the pairs  come closer to each other. That increases coupling, that tries to fix the
phases of neighboring  pairs. Therefore we expect a quantum transition to
"superconducting" phase. At the transition point, the "spin gap" vanishes to zero.
Instead, long-range phase ordering appears. That is, $x,y$ components
of the local spin density become non-zero.

To give a quantitative description, we evaluate first the phase-dependent
interaction between  pairs. It appears that this
interaction measures up $E_c$ at interskyrmion distances much larger
than screening radius $r_s$, $(r_s/\lambda)^2 = \sqrt {\pi}/ (4\sqrt 2 \tilde g)$ this is why we shall evaluate it for interpair
distances $r$ in the region $r\gg r_s,b$. We expand energy functional
up to the terms quadratic in $s_{x,y}$ and minimize the resulting expression
matching $s_{x,y}$ near the pair cores with asymptotics of $W(z)$.
That yields
\begin{equation}
E_{int}= E_J \cos ( \phi _1 - \phi_ 2), \quad E_J = g\mu _B B 
 \frac{b_0^4}{r_s^4}\exp(-\frac{r}{r_s}) (\frac{r_s}{\lambda ^2}  
\sqrt{\pi r_s r/2}).
\label{9}
\end{equation}    
 As a result, 
we have the following many-body Hamiltonian, which takes into account 
"charging" and Josephson energies:
\begin{equation}
\hat H = \sum _i E_c (-i  \frac{\partial}{\partial \phi _i} - S_0 )^2 + 
\frac{1}{2} \sum _{ij}^{\prime} E_J \cos ( \phi _i - \phi_ j),
\label{10}
\end{equation}
where $E_c$ is given by Eq.(\ref{8}),
$E_J > 0$ is given by Eq.(\ref{9}). Indexes label sites of triangular
lattice, the prime restricts sum over $j$ to the nearest 
neighbours of a certain site. 
Note that in our case the "Josephson" coupling has an antiferromagnetic sign,
 to be contrasted with Josepson arrays.
In the limit $E_J \gg E_c$ the phases become well-defined and we can neglect
"charging" term. The Hamiltonian reduces to energy functional. In the ground
state, the lattice subdivides into three sublattices.\cite{Lee} The phases of
sublattices are rotated by $2 \pi/3$ with respect to each other (Fig 2b).

We are interested in characterizing the transition.
Since the exact solution of the Hamiltonian is beyond the reach,
we find the transition point in Hartree approximation. We approximate
the exact ground-state wave function by a product of site wave functions,
\begin{equation}
\Psi(\phi_1, ..., \phi_N) =\prod_{i=1}^N \Psi_i(\phi_i),
\end{equation}
$\Psi_i$ to be determined from the minimum energy principle. The site
wave functions then satisfy the following mean-field Hamiltonian:
$$
\hat H_{MF} = \sum _i E_c (-i\frac{\partial}{\partial \phi _i} - S_0 )^2 + 
\frac{1}{4}
E_J \sum _{ij}^{\prime} [ 2n_j \exp (-i \phi _i) -n_i  n_j^* + c.c.].
$$
where we introduce the order parameter 
$$
n_i=\int_0^{2 \pi} \exp(i\phi) |\Psi_i(\phi)|^2 d \phi
$$ 
at each site. We assume that near transition point the order
parameter has the same symmetry as for $E_J \gg E_c$. The non-trivial 
solution for the order
parameter first appear at 
\begin{equation}
E_J^{0}= \frac{E_c}{3} \left \{ 1-4([S_0 + 1/2]-S_0)^2 \right \},
\label{12}
\end{equation}
this equation determining the transition point.
Making use of  Eqs.(\ref{8}, \ref{9}, \ref{12}) we find that 
the transition occurs at skyrmion density  
$n_{sky} \sim r_s^{-2} \ln^{-2} (1/\tilde g)$. To give more detailed predictions,
we plot the critical skyrmion density versus $\tilde g$ in realistic range of
this parameter. The curve exhibits sharp drops each time the energies of
two spin states match, since this favours spin fluctuations and, consequently,
facilitates phase ordering.


To observe the transition experimentally, one shall measure either spin gap
 or the local magnetization parallel to the plane, those quantities respectively vanish
and appear above transition point.
Both quantities are likely to be accessed in magnetic resonance measurement.
The experiment shall be performed at temperature 
that is lower than the typical energy scale involved, $E_c$.
Our estimations for $n_{el}=10^{16} m^{-2}$, $\tilde g= 0.025$ give $E_c/k_B \simeq 1 K$.
The fine tuning of the filling factor in the range of $0.01$ is required.
Our model does not explicitly include disorder. However we expect that
the quantum transition occurs even in disordered heterostructures, provided
the skyrmions survive. The disorder may trap the skyrmions in random potential minima,
thus preventing the formation of any regular lattice. Albeit it can not pin
the skyrmion phase, that to be pinned only by interskyrmion interactions
 with increasing concentration. This is why we expect the quantum
 transition at $n_{sky}/n_{el} \simeq \tilde g$ even for disordered arrangement 
of skyrmions. 

In conslusion, we have investigated many-skyrmion ground state at low skyrmion
concentration. We have shown that the skyrmions are bound in pairs, those
forming triangular Wigner lattice. The quantun phase transition occurs
in this lattice at certain skyrmion concentration.

The authors wish to thank G. E. W. Bauer, L. P. Kouwenhoven, M. V. Feigelman and
S. A. Kivelson for very instructive discussions  of our results.
We acknoledge financial support of
"Nederlandse Organisatie voor Wetenschappelijk Onderzoek (NWO)"
that made this work possible. A.V. Khaetskii wishes to thank the kind hospitality of
prof. G.E.W. Bauer and Dr. Yu.V. Nazarov during his visit to Delft TU.

\begin{figure}
\caption
{Binding of two skyrmions. Two skyrmion energy as a function of their
separation has a minimum at zero separation.}
\label{fig1}
\end{figure}

\begin{figure}
\caption
{Two phases of skyrmion lattice. In "insulating" phase, skyrmion phase strongly
fluctuates so that {\it x,y} spin components have zero average.
In "superconducting" phase, skyrmion phases are fixed and  antiferromagnetically arranged}
\label{fig2}
\end{figure}

\begin{figure}
\caption
{ The quantum phase transition. Critical skyrmion density is plotted
in the region of realistic $\tilde g$. Above the line, the system is in "superconducting"
state. Numbers denote the equilibrium spin of isolated skyrmion pairs at
given $\tilde g$. The crtical line drops to zero any time the spin state changes.}
\label{fig3}
\end{figure}


\begin{references}
\bibitem{Super} The skyrmion pairs are bosons so that they could
form superconducting condensate. Whatever a fascinating opportunity is that,
they do not do this since they are effectively localized by strong magnetic field.
Or, in other terms, the potential superconductivity is suppressed by the field.
\end{references}
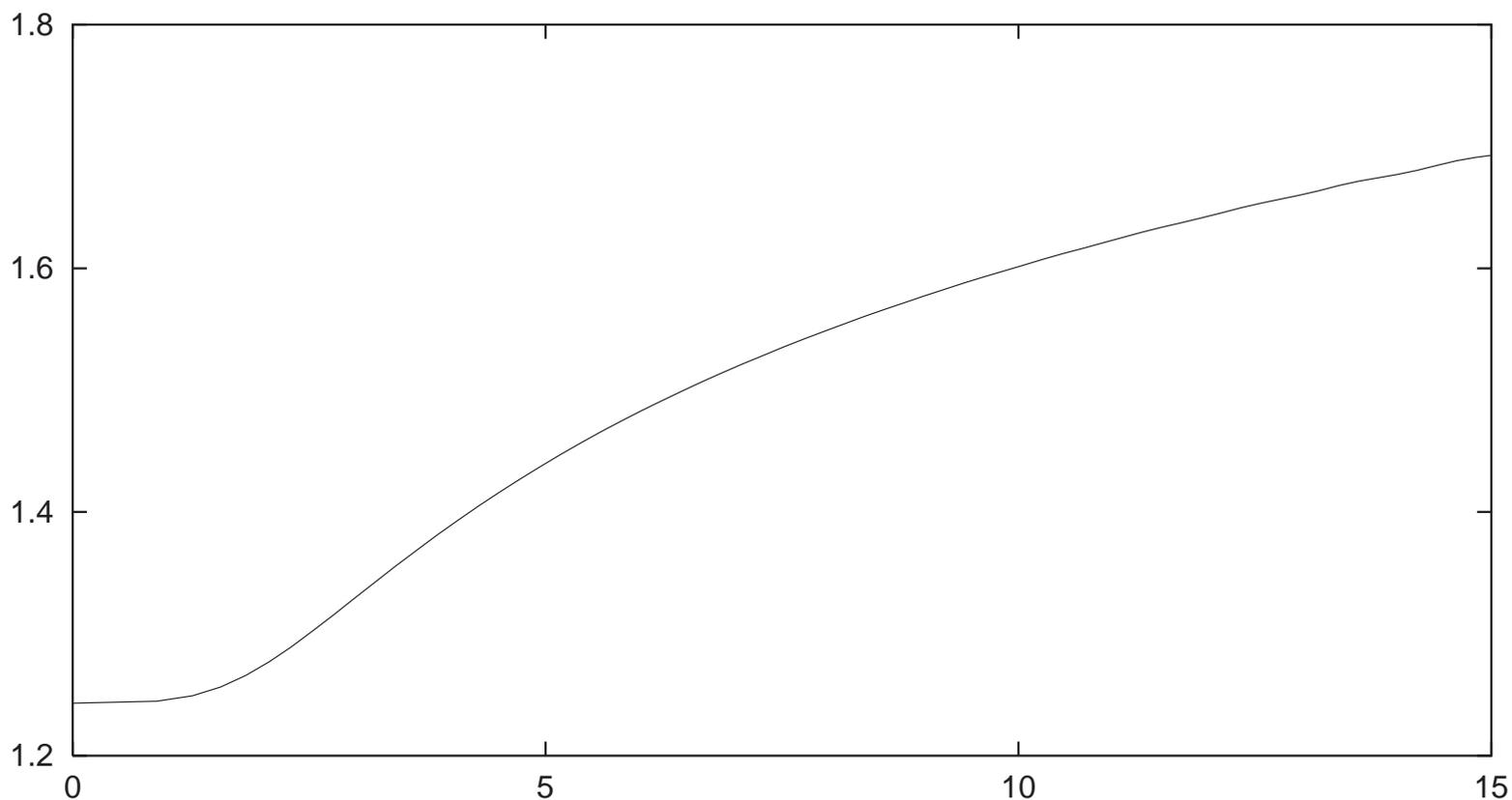
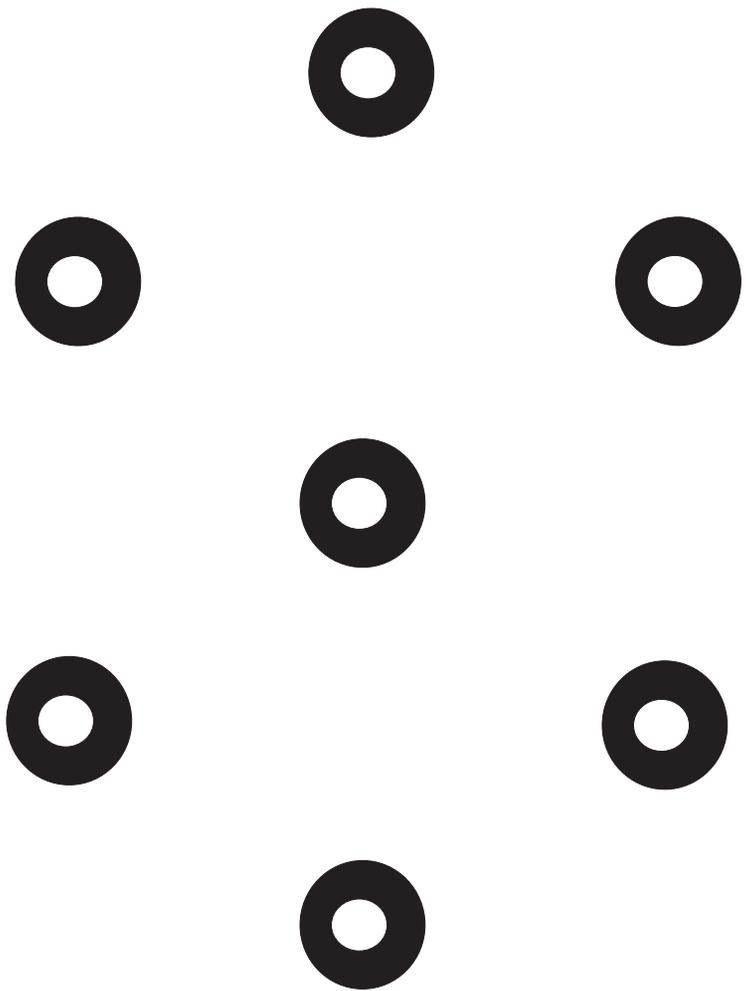
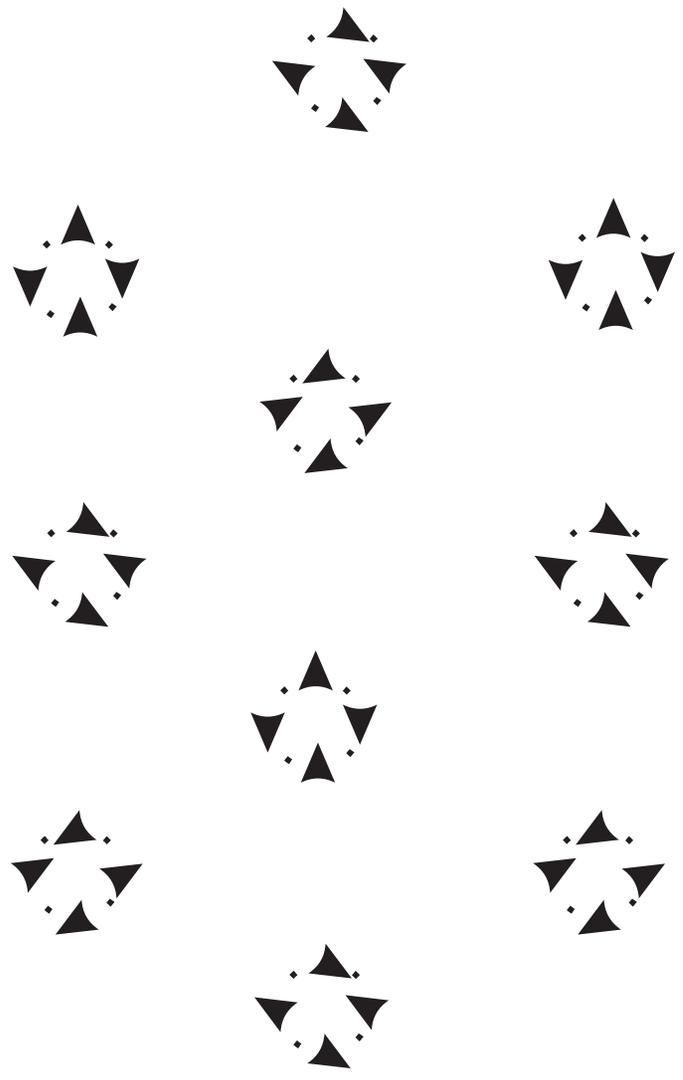
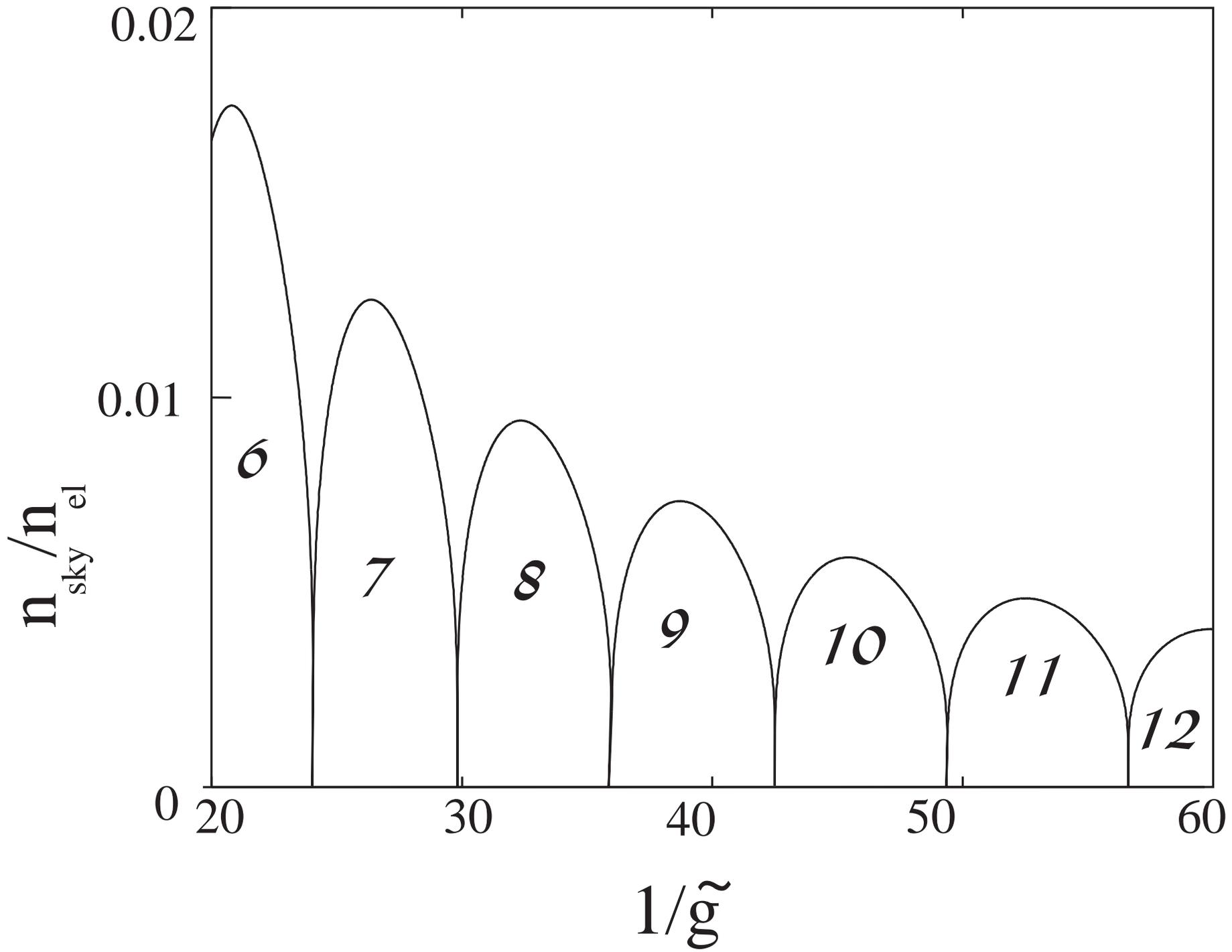
\end{document}